\documentclass[11pt]{article}
  \usepackage{amsthm}
  \usepackage{amssymb}
  \usepackage{amsmath}
  \usepackage{eucal}
  \usepackage{mathrsfs}
  \usepackage[all]{xy}
  \usepackage{bbm}

\newcommand{\M}{\mathcal{M}} 
\newcommand{\Man}{\mathfrak{Man}} 

\newcommand{\g}{\mathrm{g}} 
\newcommand{\J}{\mathrm{J}}%
\newcommand{\D}{\mathrm{D}}
\newcommand{\I}{\mathrm{I}}%
\newcommand{\C}{\mathcal{C}} 
\newcommand{\Mk}{\mathbb{M}^4} 
\newcommand{\dc}{\mathcal{O}} 
\newcommand{\K}{\mathcal{K}}  
\newcommand{\rd}{\mathcal{D}} 
\newcommand{\Kr}{\mathcal{K}_{\diamond}} 
\newcommand{\A}{\mathcal{A}} 
\newcommand{\Al}{\mathscr{A}} 
\newcommand{\St}{\mathcal{S}} 

\newcommand{\ga}{\gamma}

\newcommand{\eps}{\varepsilon}
\newcommand{\io}{\iota}
\newcommand{\al}{\alpha}
\newcommand{\la}{\lambda}

\newcommand{\Hom}{\mathrm{Hom}}
\newcommand{\f}{\mathrm{f}}
\newcommand{\sst}[1]{\scriptscriptstyle{#1}}

\newcommand {\defi}{\equiv}
\newcommand {\norm}[1]{\Vert{#1}\Vert}
\newcommand{\R}{\mathbb{R}} 
\newcommand{\Hil}{\mathcal{H}} 
 \author{Giuseppe Ruzzi \\
  \small{Dipartimento di Matematica, Universit\`a di Roma ``Tor Vergata'' }\\
     \small{Via della Ricerca Scientifica I-00133, Roma,  Italy}  \\
           \small{\texttt{ruzzi@mat.uniroma2.it}}}

\title{Punctured Haag duality in locally covariant quantum field theories}
\date{}
 
\begin{document}

  \maketitle

 \begin{abstract}
We investigate a new property of nets of local algebras over 4-dimensional
globally
hyperbolic spacetimes, called \textit{punctured Haag duality}.
This property  consists in
the usual Haag duality for the restriction of the net
to the causal complement of a point $p$ of the spacetime.
Punctured Haag duality implies Haag duality and local definiteness.
Our  main result is that, if we deal with 
a locally covariant quantum field theory in the sense of
Brunetti, Fredenhagen and Verch, then also the converse holds.
The free Klein-Gordon field
provides an example in which this property is verified.
\end{abstract}

  \theoremstyle{plain}
  \newtheorem{df}{Definition}[section]
  \newtheorem{teo}[df]{Theorem}
  \newtheorem{prop}[df]{Proposition}
  \newtheorem{cor}[df]{Corollary}
  \newtheorem{lemma}[df]{Lemma}

  \theoremstyle{definition}
  \newtheorem{oss}[df]{Remark}


\section{Introduction}
The charged sectors of a net of local observables
in a  4-dimensional globally hyperbolic spacetime
have been investigated in \cite{GLRV}. The sectors define
a $\mathrm{C}^*$-category in which,
except when there are geometrical obstructions,
the charge structure arises from a tensor product, a symmetry
and a conjugation. Geometrical obstructions occur
when the spacetime has compact Cauchy surfaces: in this situation
neither the classification of the statistics of sectors,
nor the existence of a conjugation, have been established.
Important progress in this direction has been achieved  in 
\cite{Rob}. In that paper the key assumption 
is  that, given the net of local observables 
over $\M$, its   restriction 
to the causal complement  of a point $p$ of $\M$
fulfils Haag duality. This allows the author to classify the statistics 
of sectors, and to provide several results  towards the 
proof of the existence of a conjugation.
The  property assumed in \cite{Rob}, which we call
punctured Haag duality,
is the subject of the present paper. 
We will demonstrate that a net satisfying punctured
Haag duality  is locally definite and fulfils Haag duality. 
Furthermore,  if we deal with a locally covariant quantum 
field theory \cite{BFV},  then we are able to show that punctured Haag 
duality is equivalent 
to Haag duality plus local definiteness.
This will allow us to provide
an example, the free Klein-Gordon field, satisfying punctured
Haag duality.\\[3pt]
\indent According to \cite{BFV}, once a locally covariant quantum
field theory  is given,
to any  4-dimensional globally hyperbolic spacetime $\M$,
there corresponds a net of local algebras
$\Al_{\K(\M)}$ satisfying the Haag-Kastler axioms
\cite{HK}. Namely, $\Al_{\K(\M)}$ is an inclusion preserving map
$\K(\mathcal{M})\ni\dc\longrightarrow\A(\dc)\subset\Al(\M)$
assigning to each element $\dc$ of a collection
$\K(\M)$ of subregions of $\M$
a $\mathrm{C}^*$-subalgebra $\A(\dc)$ of a $\mathrm{C}^*$-algebra $\Al(\M)$,
and satisfying
\[
\qquad \dc_1\perp\dc   \ \ \Rightarrow   \ \ [\A(\dc_1),\A(\dc)]=0
\qquad locality.
\]
In words locality means that if $\dc_1$ is causally disjoint from $\dc$,
then the algebras $\A(\dc_1)$, $\A(\dc)$ commute elementwise.
$\A(\dc)$ is the algebra generated by all the observables measurable
within the region $\dc$. The main new feature with respect to the
Haag-Kastler framework is that if
$\psi$ is an isometric embedding  from $\M$
into another 4-dimensional globally hyperbolic spacetime $\M_1$,
then there is a $\mathrm{C}^*$-morphism $\alpha_{\psi}$ from
$\Al(\M)$ into $\Al(\M_1)$ which maps isomorphically
$\A(\dc)$ onto $\A(\psi(\dc))$
for each $\dc\in\K(\M)$ (see \cite{BFV} or
Section \ref{Ca} for details).\\[3pt]
\indent Now, turning  to the subject of the present paper,
given a state $\omega$ of $\Al(\M)$
and denoting by $\pi$ the GNS representation of $\omega$,
we  say that $(\A_{\K(\M)},\omega)$ satisfies
\textit{punctured Haag duality} if for any point
$p$ of $\M$ the following identity
\begin{equation}
\label{I:3}
\pi(\A(\rd_1))'' =  \ \cap \ \{ \ \ \pi(\A(\rd))' \ \ |\rd\in\Kr(\M), \
  \rd\perp (\rd_1\cup\{p\})\}
\end{equation}
is verified for any $\rd_1\in\Kr(\M)$ such that
$\rd_1\perp\{p\}$, where $\Kr(\M)$ is  the subcollection
of $\K(\M)$ formed by
the regular diamonds of $\M$ (see Section \ref{Ba}).
In the Haag-Kastler framework, punctured Haag duality
implies Haag duality and
local definiteness (Proposition \ref{Cb:1}). 
The main result of the
present paper is that, in the setting of a  locally
covariant quantum field theory, punctured Haag duality is equivalent 
to Haag duality plus local definiteness 
(Theorem \ref{Cc:4} and Corollary \ref{Cc:5}).\\[3pt] 
\indent The basic idea of the proof of Theorem \ref{Cc:4} is the
following. The condition $\rd\perp\{p\}$ means that the closure 
$\overline{\rd}$
of $\rd$ is contained in the open set $\M\setminus\J(p)$.
The spacetime $\M_p\defi\M\setminus\J(p)$
is globally hyperbolic, hence, there is a net of local algebras
$\Al_{\K(\M_p)}$ associated with $\M_p$. Now, as
the injection $\io_p$ of $\M_p$ into $\M$ is an isometric
embedding, local covariance allows us
to identify the net $\Al_{\K(\M_p)}$ with
the restriction of $\Al_{\K(\M)}$ to $\M\setminus\J(p)$, and the
algebra $\Al(\M_p)$ with a subalgebra $\Al(\M)$.
Then,  if  $\omega$ is a  state of $\Al(\M)$,
punctured Haag duality for $(\Al_{\K(\M)},\omega)$
seems to be equivalent to  Haag duality for
$(\Al_{\K(\M_p)},\omega|_{\Al(\M_p)})$.
This is actually true, but, there is a subtle point
that has to be carefully considered:
the regular diamonds of $\M_p$ might not be regular
diamonds of $\M$. However, we will be able to circumvent this
problem by means of the following result:
the set $\Kr(\M_p)$ and the set
$\{\rd\in\Kr(\M) \ | \overline{\rd}\subset \M\setminus\J(p)\}$
have a common ``dense''  subset (Proposition~\ref{Bb:7}).\\[3pt]
\indent As an easy consequence of this result,
we will show that the free Klein-Gordon field over
a 4-dimensional globally hyperbolic spacetime,
in the representation associated with a pure
quasi-free state satisfying  the
microlocal  spectrum condition, fulfils punctured Haag duality
(Proposition~\ref{Cd:1}).  The same holds
for pure adiabatic vacuum states of order $N>\frac{5}{2}$ (Remark \ref{Cd:2}).
\section{Preliminaries on spacetime geometry}
\label{A}
We recall some basics on the causal
structure of spacetimes and establish our notation.
Standard references for this topic are \cite{One, EH, Wal}.\\[3pt]
\textbf{Spacetimes} A \textit{spacetime} $\M$ consists in
a Hausdorff, paracompact,  smooth, oriented
4-dimensional manifold  $\M$ endowed
with  a smooth metric $\g$  with signature $(-,+,+,+)$, and with
a time-orientation, that is a  smooth vector field $v$ satisfying
$\g_p(v_p,v_p) < 0$ for each $p\in\M$.
(Throughout this paper smooth means $C^\infty$). \\
\indent A  curve $\ga$ in $\M$ is a continuous,  piecewise smooth, 
regular function $\ga:I \longrightarrow \M$,  
where $I$ is a connected subset of $\R$ with nonempty interior. It
is  called timelike, lightlike, spacelike 
if respectively $\g(\dot{\ga},\dot{\ga})<0$, $=0$, $> 0$
all along $\ga$, where $\dot{\ga}=\frac{d\ga}{dt}$.
Assume now that $\ga$ is \textit{causal}, i.e.
a  nonspacelike curve; we can classify it according to the 
time-orientation $v$ as future-directed (f-d) or 
past-directed (p-d) if respectively 
$\g(\dot{\ga}, v) < 0, >0$ all along $\ga$. When  $\ga$ is f-d and 
there exists $\lim_{t\rightarrow\sup I}\ga(t)$ 
($\lim_{t\rightarrow\inf I}\ga(t)$), then it is said to have 
a future (past) endpoint. In the negative case, it is said to be 
future (past) endless;  $\ga$ is said to be endless if none of them
exist. Analogous  definitions are assumed for p-d causal curves.\\
\indent The \textit{chronological future} $\I^+(S)$,
the \textit{causal future} $\J^+(S)$
and the \textit{future domain of dependence } $\D^+(S)$
of a subset $S\subset\M$ are defined as:
\begin{align*}
&\I^+(S)  \defi  \{ x\in\M \ | \ \mbox{there is a f-d timelike curve
    from $S$ to $x$ } \}; \\
&\J^+(S)  \defi   S \cup \{ x\in\M \ | \ \mbox{there is a f-d  causal
    curve from $S$ to $x$ } \}; \\
&\D^+(S) \defi   \{ x\in\M \ | \ \mbox{any p-d endless causal
    curve through $x$ meets $S$ } \}.
\end{align*}
These definitions
have a dual in which ``future'' is replaced by ``past''
and the $+$ by $-$.  So,
we define $\I(S) \defi \I^+(S) \cup  \I^-(S)$,
$\J(S)\defi\J^+(S) \cup  \J^-(S)$ and $\D(S) \defi\D^+(S)\cup
\D^-(S)$.  We recall that
$\I^+(S)$ is an open set, and that
$\overline{\J^+(S)} = \overline{\I^+(S)}$ and $(\J^+(S))^o = \I^+(S)$.\\
\indent Two sets $S,V\subset \M$ are \textit{causally disjoint},
$S\perp V$, if $\overline{S}\subseteq\M\setminus\J(\overline{V})$
or, equivalently, if $\overline{V}\subseteq\M\setminus\J(\overline{S})$.
A set $S$ is \textit{achronal} if
$S\cap \I(S) = \emptyset$; it is \textit{acausal}
if $\{p\}\perp \{q\}$ for each pair $p,q\in S$.
A (\textit{acausal}) \textit{Cauchy surface}
$\C$ of $\M$ is an achronal (acausal) set verifying $\D(\C)=\M$.
Any Cauchy surface is a closed, connected, Lipschitz hypersurface of
$\M$. A \textit{spacelike} Cauchy surface is a smooth Cauchy surface
whose tangent space is everywhere spacelike.  Any spacelike Cauchy
surface is acausal.\\[3pt]
\textbf{Global hyperbolicity} A spacetime $\M$ is
\textit{strongly causal} if for each
point $p$ the following condition
holds: any neighbourhood $U$ of $p$ contains
a neighbourhood $V$ of $p$ such that
for each $q_1,q_2\in V$ the set
$\J^+(q_1)\cap \J^-(q_2)$ is either empty or contained in $V$.
$\M$ is \textit{globally hyperbolic}
if it admits a Cauchy surface or, equivalently, if it is strongly
causal and for each pair of points $p_1$, $p_2$  the set
$\J^+(p_1)\cap \J^-(p_2)$ is empty or compact.
Assume that $\M$ is globally hyperbolic. Then $\M$ can be foliated by
spacelike
Cauchy surfaces \cite{Die}. Namely, there is 3-dimensional smooth manifold
$\Sigma$ and a diffeomorphism $F:\R \times \Sigma \longrightarrow \M$
such that:  for each $t\in\R$ the set $\C_t\defi \{F(t,y) \ | \ y\in\Sigma\}$
is a spacelike Cauchy surface of $\M$; for each $y\in\Sigma$,
the curve $t\in\R\longrightarrow F(t,y)\in\M$ is  a f-d (by convention)
endless timelike curve. For any relatively compact $S\subset\M$
the following properties hold: \textbf{1.}~$\J^+(\overline{S})=\overline{\J^+(S)}$;
\textbf{2.}
$\D^+(\overline{S})$ is compact; \textbf{3.} for each
Cauchy surface $\C$ the set $\J^+(\overline{S})\cap \C$
is either empty or compact; \textbf{4.}
$\overline{\J^+(S\cup \{p\})} =  \J^+(\overline{S}\cup \{p\})$
for any  $p\in\M$ such that $\overline{S}\cap \{p\}=\emptyset$.\\[3pt]
\textbf{The category $\boldsymbol{\mathfrak{Man}}$} \cite{BFV}
Let $\M$ and $\M_1$ be  globally hyperbolic spacetimes
with metric $\g$ and $\g_1$ respectively.
A smooth function $\psi$ from $\M_1$ into $\M$ is called an
\textit{isometric embedding}  if $\psi:\M_1\longrightarrow \psi(\M_1)$
is a diffeomorphism and $\psi_*\g_1 = \g|_{\psi(\M_1)}$.
The category $\mathfrak{Man}$
is the category whose objects are the
4-dimensional globally  hyperbolic spacetimes; the arrows
$\Hom(\M_1,\M)$ are the isometric embeddings
$\psi:\M_1\longrightarrow\M$ preserving the orientation
and the time-orientation of the embedded spacetime, and that satisfy
the property
\[
\forall p,q\in\psi(\M_1), \ \J^+(p)\cap \J^-(q) \mbox{ is either empty or
contained  in } \psi(\M_1).
\]
The composition law between two arrows $\psi$ and $\phi$, denoted by
$\psi\circ \phi$, is given by the usual composition between
smooth functions; the identity
arrow  $id_{\M}$ is the identity  function of $\M$.
\section{Causal excisions  and regular diamonds}
\label{B}
We present two index sets for nets of local algebras over a globally
hyperbolic spacetime $\M$: the set $\K(\M)$ and the set of the regular
diamonds $\Kr(\M)$. Furthermore, we introduce the spacetime
$\M_p$, the  causal excision of $p\in\M$, and study
how $\K(\M_p)$ and $\Kr(\M_p)$ are embedded in $\M$.
\subsection{The sets $\K(\M)$ and $\Kr(\M)$}
\label{Ba}
Let us consider a globally hyperbolic spacetime $\M$ with metric $\g$.
The set $\K(\M)$ \cite{BFV}  is defined as
the collection of the open sets $\dc$ of $\M$ which are
relatively compact,  open, connected  and enjoy the following
property: for each pair  $p,q\in \dc$  the set
      $\J^+(p)\cap \J^-(q)$  is either  empty or contained in
$\dc$.
The importance of $\K(\M)$
for the locally covariant quantum field theories derives from the following
properties that can be easily checked:
\begin{align}
\label{Ba:1}
\dc\in\K(\M)  & \ \ \ \Rightarrow  \ \ \
\dc\in\mathfrak{Man}\\
\label{Ba:2}
\psi\in\Hom(\M_1,\M)  & \ \
 \ \Rightarrow \ \ \ \psi(\K(\M_1))\subseteq\K(\M),
\end{align}
where for $\dc\in\mathfrak{Man}$ we mean that $\dc$ with the metric
$\g|_{_{\dc}}$ and with induced orientation and time orientation is
globally hyperbolic.
Property (\ref{Ba:1}) allows one to associate a net
of local algebras with $\M$; property (\ref{Ba:2}) makes
the algebras associated with isometric regions
of different spacetimes isomorphic. $\K(\M)$ however is a too big
index set for  studying properties of the net like
Haag duality and punctured Haag duality (see Section \ref{Cb}).
For this purpose  the set of the regular diamonds of $\M$
is well suited \cite{GLRV, Ve1}.
Given a smooth manifold $\mathcal{N}$, let
\begin{itemize}
\item[$\mathfrak{G}(\mathcal{N})$] $\defi \{ \  G\subset \mathcal{N}
       \ | \ G$
      and  $\mathcal{N}\setminus \overline{G}$
      are nonempty and open,
     $\overline{G}$  is compact
      and contractible  to a point in $G$;
     $\partial G$  is a two-sided, locally flat
     embedded $C^0$-submanifold of $\mathcal{N}$
     having finitely many connected
     components, and in each connected component there are
      points near to which $\partial G$
     is  $C^\infty$-embedded $\}$.
\end{itemize}
Then,  a \textit{regular diamond} $\rd$ is an open subset of $\M$
of the form $\rd=(\D(G))^o$ where
$G\in\mathfrak{G}(\C)$ for some \textit{spacelike} Cauchy surface
$\C$; $\rd$ is said to be  \textit{based} on $\C$, while  $G$ is called the
\textit{base} of $\rd$.
We denote
 by $\Kr(\M)$ the set of the
regular diamonds of $\M$, and by  $\Kr(\M,\C)$
those elements of $\Kr(\M)$ which are based on  the spacelike Cauchy
surface $\mathcal{\C}$.\\[3pt]
\indent The set $\Kr(\M)$ is a base of the topology of $\M$ and
$\Kr(\M)\subset\K(\M)$. Moreover, let us consider  $\rd\in\Kr(\M,\C)$ 
and an open set $U$
such that $\overline{\rd}\subset U$. Then there exists (\cite[Lemma 3]{Ve1})
$\rd_1\in\Kr(\M,\C)$ such that 
\begin{equation}
\label{Ba:3}
\overline{\rd}\subset \rd_1  \qquad 
 \overline{\rd}_1\subset (\D(U\cap \C))^o. 
\end{equation}
Now, notice that
\[
\psi\in \Hom(\M_1,\M) \ \ \ \Rightarrow \ \ \ 
  \psi(\Kr(\M_1))\subset \K(\M),
\]
but $\psi(\Kr(\M_1))$ might \textit{not} be 
contained in $\Kr(\M)$. To see this, consider a spacelike 
Cauchy surface $\C_1$  of $\M_1$ and notice that  
$\psi(\C_1)$ is a spacelike hypersurface of $\M$.
If $\rd_1$ is a regular diamond of $\M_1$ of the form $(\D(G_1))^o$, 
where $G_1\in\mathfrak{G}(\C_1)$, then  
\[
\psi((\D(G_1))^o) = (\D(\psi(G_1)))^o \ \ \ \mbox{ and } \ \ \ 
	\psi(G_1)\in\mathfrak{G}(\psi(\C_1))
\]
However, in general a spacelike 
Cauchy surface $\C$ of
$\M$ such that $\psi(\C_1)\subseteq \C$ does not exist
(see for instance Remark \ref{Bb:4}). More in general, we do not know 
whether $\psi(G_1)\in\mathfrak{G}(\tilde{\C})$ 
for some  spacelike Cauchy surface $\tilde{\C}$ of $\M$.
So, we have no way to conclude that $\psi(\rd)\in\Kr(\M)$.
\subsection{Causal excisions}
\label{Bb}
Consider a globally hyperbolic spacetime $\M$, with metric $\g$,
and  a point $p$ of
$\M$. As the set
$\M\setminus \J(p)$ is open and connected, the manifold
\begin{equation}
\label{Bb:1}
\M_p\defi\M\setminus \J(p) \mbox{ endowed with the metric }
\g_p\defi \g|_{\M\setminus \J(p)},
\end{equation}
and with the induced orientation and time-orientation, is a spacetime.
$\M_p$ inherits from $\M$ the strong causality
condition because this property is stable under restriction to open subsets.
Moreover,  for each pair $p_1,p_2\in\M\setminus \J(p)$
the set $\J^+(p_1)\cap\J^-(p_2)$ is either empty or compact and
contained in $\M\setminus \J(p)$ (the compactness follows from the
global hyperbolicity of $\M$). Hence $\M_p$ is globally
hyperbolic, thus $\M_p\in\mathfrak{Man}$.
We call $\M_p$  \textit{the causal excision of $p$}.
Let us now denote by $\io_p$ the injection
of $\M_p$ into $\M$, that is
\[
\M\setminus\J(p)\ni q\longrightarrow \io_p(q)=q\in\M.
\]
Clearly, $\io_p\in\Hom(\M_p,\M)$.
\begin{prop}
\label{Bb:3}
Given a globally hyperbolic spacetime $\M$,
let $\M_p$ be the causal excision of $p\in\M$.
If $\C$ is an acausal (spacelike) Cauchy surface of $\M$ that meets $p$,
then $\C\setminus \{p\}$ is an acausal (spacelike) Cauchy surface of $\M_p$.
Conversely, if $\C_p$ is an acausal Cauchy surface of $\M_p$,
then $\C_p\cup \{p\}$ is an acausal Cauchy surface of $\M$.
\end{prop}
\begin{proof}
$(\Rightarrow)$ $\C\setminus \{p\}$ is an acausal set
of  $\M_p$ and $\D^+(\C\setminus \{p\})\subseteq \D^+(\C)\setminus \J(p)$.
Conversely, if $q\in  \D^+(\mathcal{C}) \setminus \J(p)$ each
p-d  endless causal curve  through $q$
meets $\mathcal{C}$ but not $p$. Hence $\D^+(\mathcal{C}\setminus \{p\})
\supseteq \D^+(\mathcal{C}) \setminus \J(p)$. This implies
$\D^+(\mathcal{C}\setminus \{p\}) =  \D^+(\mathcal{C}) \setminus \J(p)$
and the dual equality
$\D^-(\mathcal{C}\setminus \{p\})=  \D^-(\mathcal{C}) \setminus
\J(p)$.
Therefore
$\D(\mathcal{C}\setminus \{p\})$ $ =
  \D^+(\mathcal{C})\setminus \J(p)  \  \cup \
  \D^-(\mathcal{C})\setminus \J(p)$
 $=  \left(\D^+(\mathcal{C})\cup \D^-(\mathcal{C})\right)
   \setminus \J(p)$  $=  \M\setminus \J(p)$.
Clearly if $\C$ is spacelike, then $\C\setminus \{p\}$ is spacelike.
This completes the proof. $(\Leftarrow)$ Set $\C\defi \C_p\cup \{p\}$ and
notice
that $\M\setminus \J(p)=\D(\mathcal{C}_p)\subseteq \D(\C)$.
Now, given $q\in \J^+(p)$ let
us consider a p-d endless causal curve $\ga$ through $q$ that
does not meet $p$. Because of global hyperbolicity
$\ga$ leaves $\J^+(p)$.
Each connected component of $\ga\cap (\M\setminus \J(p))$ is
a f-d endless causal curve of $\M_p$, therefore
it meets $\C_p$. This entails
$\J(p)\subset\D( \C )$ and that $\D(\C)=\M$. In order to prove that
$\C$ is acausal,  assume that there exists
a f-d causal curve  $\ga:[0,1]\longrightarrow \M$ that joins two
points $q_1$ and $q_2$ lying on $\C$.
If one of the two points is $p$, then $\ga(t)\in \J(p)$ for each
$t\in[0,1]$, and this leads to a contradiction. If $q_1,q_2\ne p$,
and $\ga \cap \J(p)=\emptyset$, then $\ga$ would be a f-d causal curve
of $\M_p$ joining two points of $\C_p$, and this leads to a contradiction.
The same happens in the case where
$q_1,q_2\ne p$ and $\ga\cap \J(p)\ne \emptyset$.
In fact, if $\ga(t_1)\in \J^+(p)$,  then
$\ga(t)\in \J^+(p)$ for each $t\geq t_1$. Analogously,
if $\ga(t_1)\in \J^-(p)$,  then $\ga(t)\in \J^-(p)$ for each $t\leq t_1$.
\end{proof}
\begin{oss}
\label{Bb:4}
Let $\C_p$ be a spacelike Cauchy surface  of $\M_p$. Because
of Proposition \ref{Bb:3},
$\C_p\cup\{p\}$ is  an acausal Cauchy surface
of $\M$. However, it is not smooth in general.
Consider for instance  the Minkowski space $\Mk$. Let
\begin{align*}
\Mk_{\underline{o}}& \defi \{ (t,\vec{x})\in\Mk \ | \ -t^2 +
\left<\vec{x},\vec{x}\right>  >  0\},\\
\C_{\underline{o}} & \defi \{(t,\vec{x})\in\Mk \ | \
- 4\cdot t^2 +  \left<\vec{x},\vec{x}\right> = 0 , \ t >0\},
\end{align*}
where $\left<,\right>$ denotes the canonical scalar product of $\mathbb{R}^3$.
$\Mk_{\underline{o}}$ is the causal excision of
$\underline{o}=(0,0,0,0)$,
and  $\C_{\underline{o}}$ is  a spacelike Cauchy surface of
$\Mk_{\underline{o}}$. Clearly,
$\mathcal{C}_{\underline{o}}\cup \{\underline{o}\}$ is a nonsmooth
hypersurface of $\Mk$.
\end{oss}
We now turn to study  how the injection $\io_p$ embeds
$\K(\M_p)$ and $\Kr(\M_p)$ into  $\M$. Concerning $\K(\M_p)$ one
can easily prove that
\begin{equation}
\label{Bb:5}
 \K(\M_p) = \{ \dc\in \K(\M) \ | \ \dc \perp \{p\} \  \}.
\end{equation}
As for regular diamonds, in general we do not know whether
$\Kr(\M_p)\subset \Kr(\M)$
(see the observation made in the previous
section). However, the set 
\begin{equation}
\Kr(\M_p\wedge\M)\defi \Kr(\M_p)\cap \Kr(\M)
\end{equation}
of the regular diamonds shared by $\M_p$ and $\M$ is not empty. 
In fact, notice that  
if $\C$ is a spacelike Cauchy surface of $\M$ that meets
$p$, then by Proposition \ref{Bb:3}
$\C_p\defi \C\setminus \{p\}$ is a spacelike
Cauchy surface of $\M_p$. Since $\mathfrak{G}(\C_p)\subset\mathfrak{G}(\C)$,
we conclude that
\begin{equation}
\label{Bb:6}
 \Kr(\M_p,\C_p) = \{\rd\in\Kr(\M,\C) \ | \ \rd \perp \{p\} \  \}
  \ \subset \ \Kr(\M_p\wedge\M).
\end{equation}
Furthermore,  $\Kr(\M_p\wedge\M)$ 
is a ``dense''  subset of both  $\Kr(\M_p)$ and 
$\{\rd\in\Kr(\M) \ | \ \rd \perp \{p\}\}$, as the following 
proposition shows. 
\begin{prop}
\label{Bb:7}
For each pair $\rd,\rd_1\in\Kr(\M_p)$ such that
$\overline{\rd}\subset\rd_1$, there exists
$\rd_o\in \Kr(\M_p\wedge\M)$ such that
$\overline{\rd}\subset \rd_o$,  $\overline{\rd}_o\subset\rd_1$.
The same result holds true for each
$\rd,\rd_1\in\{\rd\in\Kr(\M) \ | \ \rd \perp \{p\} \}$
such that
$\overline{\rd}\subset\rd_1$.
\end{prop}
\begin{proof}
The proof follows from Propositions \ref{ApA:3}, \ref{ApA:4}.
\end{proof}
\section{Punctured Haag duality}
\label{C}
This section is devoted to the investigation of punctured Haag duality.
We will start by recalling the axioms of a locally covariant quantum field
theory. Afterwards, we will show necessary and sufficient conditions
for punctured Haag duality, both in the Haag-Kastler framework and 
in the setting of the locally covariant quantum field theories.
Finally, we will apply these
results to the theory of the free Klein-Gordon field.
\subsection{Locally covariant quantum field theories}
\label{Ca}
The locally covariant quantum field theory is a categorical
approach to the theory of quantum fields incorporating
the covariance principle of general relativity \cite{BFV}.
In order to introduce the axioms of
the theory, we give a preliminary definition.
Let us denote by $\mathfrak{Alg}$ the category whose objects
are unital $\mathrm{C}^*$-algebras and whose
arrows  $\Hom(\A_1,\A_2)$ are the  unit-preserving injective
$\mathrm{C}^*$-morphisms from $\A_1$ into $\A_2$.
The composition law between the arrows $\al_1$ and $\al_2$,
denoted by $\al_1\circ\al_2$,  is given by the usual
composition between $\mathrm{C}^*$-morphisms;  the unit
arrow  $id_\A$ of $\Hom(\A,\A)$ is the identity morphism of $\A$. \\[3pt]
\indent A \textbf{locally covariant quantum field theory} is a
covariant functor $\Al$
from the category $\mathfrak{Man}$ (see Section \ref{A})  into
the category $\mathfrak{Alg}$, that is,
a  diagram
\[
\xymatrix{
 \M_1 \ar[d]_\Al \ar[r]^{\psi} & \M_2 \ar[d]^\Al\\
\Al(\M_1)  \ar[r]_{\alpha_\psi} &\Al(\M_2),}
\]
where $\alpha_{\psi}\defi \Al(\psi)$, such that
$\al_{id_{\M}} = id_{\Al(\M)}$, and
$\al_\phi\circ \al_\psi =\al_{\phi\circ \psi}$
for each $\psi\in \Hom(\M_1,\M)$ and
$\phi\in \Hom(\M,\M_2)$.
The functor $\Al$ is said to be \textbf{causal}
if, given  $\psi_i\in \Hom(\M_i,\M)$
for $i=1,2$,
\[
\psi_1(\M_1) \perp \psi_2(\M_2) \ \Rightarrow \
\left[ \alpha_{\psi_1}(\Al(\M_1), \alpha_{\psi_2}(\Al(\M_2))\right] = 0,
\]
where $\psi_1(\M_1)\perp\psi_2(\M_2)$ means that $\psi_1(\M_1)$ and
$\psi_2(\M_2)$ are causally
disjoint in $\M$. From now on $\Al$ will denote
a causal locally covariant quantum field theory.\\[3pt]
\indent We now turn to the notion of a state space
of $\Al$. To this aim,  let
$\mathfrak{Sts}$  be the category
whose objects are the state spaces $\St(\A)$
of unital $\mathrm{C}^*$-algebras  $\A$, namely
$\St(\A)$ is a subset of the states of $\A$ closed under finite
convex combinations and operations $\omega(\cdot)\rightarrow \omega(A^*\cdot
A)/\omega(A^*A)$ for $A\in\A$. The arrows  between two objects
$\St(\A)$ and $\St'(\A')$ are the positive
maps $\ga^*: \St(\A)\longrightarrow \St'(\A')$,
arising as dual maps
of injective morphisms of $\mathrm{C}^*$-algebras
$\ga:\A'\longrightarrow \A$, by
$\ga^*\omega(A) \defi \omega(\ga(A))$  for each $A\in\A$.
The composition law between two arrows, as the definition of the identity
arrow of an object, are obvious. A \textbf{state space} for
$\Al$ is a contravariant functor
$\mathcal{S}$ between $\mathfrak{Man}$ and $\mathfrak{Sts}$, that is,
a diagram
\[
\xymatrix{
\M_1  \ar[d]_{\mathcal{S}} \ar[r]^{\psi} &
                         \M_2 \ar[d]^{\mathcal{S}}\\
\mathcal{S}(\M_1) & \ar[l]_{\alpha^*_\psi} \mathcal{S}(\M_2),
}
\]
where $\mathcal{S}(\M_1)$ is a state space of the algebra
$\Al(\M_1)$, such that
$\alpha^*_{\io_{\M}}= id_{\St(\M)}$, and
$\al^*_\psi\circ \al^*_\phi =\al^*_{\phi\circ\psi}$
for each $\psi\in \Hom(\M_1,\M)$ and
$\phi\in \Hom(\M,\M_2)$.\\[3pt]
\indent In conclusion let us see
how a net of local algebras over $\M\in\mathfrak{Man}$
can be recovered by a locally covariant quantum
field theory $\Al$. For this purpose,
recall that  by (\ref{Ba:1}) any  $\dc\in \K(\M)$, considered as a spacetime
with the metric $\g|_{\dc}$,
belongs to $\mathfrak{Man}$.
The injection $\io_{\M,\dc}$ of
$\dc$ into $\M$ is an element of
$\Hom(\dc,\M)$  because of the definition of $\K(\M)$.
Then, by using  $\al_{\io_{\M,\dc}}\in \Hom(\Al(\dc), \Al(\M))$
to define $\A(\dc)\defi \al_{\io_{\M,\dc}}\left(\Al(\dc)\right)$,
it turns out that the correspondence $\Al_{\K(\M)}$, defined as
\begin{equation}
\label{Ca:0}
\K(\M)\ni\dc\longrightarrow \A(\dc)\subset\Al(\M),
\end{equation}
is a net of local algebras satisfying the Haag-Kastler axioms.
As for the local covariance of the theory, let $\M_1\in\Man$ with
the metric $\g_1$. Notice that if
$\psi\in\Hom(\M,\M_1)$, then $\psi(\dc)\in \K(\M_1)$
for each $\dc\in\K(\M)$. Since
$\io^{-1}_{\M_1,\psi(\dc)}\circ\psi\circ \io_{\M,\dc}$ is an isometry
from the spacetime $\dc$ into the spacetime $\psi(\dc)$~---~the latter
equipped with the metric $\g_1|_{\psi(\dc)}$~---~one has that
\begin{equation}
\label{Ca:1}
\alpha_{\psi}:\A(\dc)\subset \Al(\M)\longrightarrow
               \A(\psi(\dc))\subset \Al(\M_1)
\end{equation}
is an $\mathrm{C}^*$-isomorphism.
\subsection{Punctured Haag duality in the Haag-Kastler framework}
\label{Cb}
We investigate punctured Haag duality (\ref{I:3})
in the Haag-Kastler framework. 
This means that we will study punctured Haag duality 
on the net of local algebras $\Al_{\K(\M)}$, associated with a spacetime 
$\M\in\Man$ by (\ref{Ca:0}), without making  use of the local
covariance (\ref{Ca:1}). In this framework,  we will obtain two 
necessary conditions for punctured Haag duality.\\[3pt] 
\indent To begin with, let $\omega$ be a state of the algebra 
$\Al(\M)$ and let $\pi$ be the GNS representation associated with $\omega$. 
We recall that 
$(\Al_{\K(\M)},\omega)$ is said to be \textit{locally definite}
if for each $p\in\M$,
\[
\mathbb{C}\cdot \mathbbm{1}  = \cap\{ \pi(\A(\rd))'' \ | \ 
   p\in \rd\in \Kr(\M) \};
\]
it said to satisfy  \textit{Haag duality} if for each $\rd_1\in\Kr(\M)$,
\[
\pi(\A(\rd_1))''  = \cap \{\pi(\A(\rd))' \ | \rd\in\Kr(\M), \
\rd \perp \rd_1\}.
\]
These properties
(and also  punctured Haag duality)
are defined in terms
of the local algebras associated with regular diamonds.
The reason is that Haag duality has been proved,
in models of quantum fields, only when the
local algebras are defined in regions of the spacetime like
the regular diamonds \cite{LR, Ve,JS}.\\
\indent Punctured Haag duality  is strongly  related to
Haag duality and local definiteness, as the following proposition shows.
\begin{prop}
\label{Cb:1}
Assume that $(\Al_{\K(\M)},\omega)$ 
satisfies punctured Haag duality. Then, 
 $(\Al_{\K(\M)},\omega)$ satisfies Haag duality. Furthermore, 
if $\omega$ is pure and 
\begin{equation}
\label{Cb:0}
\pi(\Al(\M))\subseteq \left( \ \cup\{\pi(\A(\dc))''| \ \dc\in\K(\M), 
\ \M\setminus\overline{\dc}\ne\emptyset\} \ \right)'',
\end{equation} 
where $\pi$ is  the GNS
representation  of $\omega$,  then $(\Al_{\K(\M)},\omega)$ is locally definite.
\end{prop}
\begin{proof}
Obviously $(\Al_{\K(\M)},\omega)$ fulfils Haag duality.
As for the proof of local definiteness,
let us define
$\mathfrak{R}_p\defi \cap \{\pi(\A(\rd))'' \ | p\in \rd\in\Kr(\M) \}$,
for  $p\in\M$.
Notice that,  as $\Kr(\M)$ contains a neighbourhoods basis for each
point $p$ of $\M$, locality entails that
$\mathfrak{R}_p\subset\pi(\A(\rd))'$
for each $\rd\in\Kr(\M), \ \rd\perp\{p\}$.
Combining this with punctured Haag duality we have that
$\mathfrak{R}_p\subset\pi(\A(\rd))''$, therefore
\[
\qquad \mathfrak{R}_p\subset \pi(\A(\rd))''\cap \pi(\A(\rd))'
\qquad \rd\in\Kr(\M), \ \rd\perp\{p\}. \qquad (*)
\]
Now, let $p_1\in\M$ be such that $\{p\}\perp\{p_1\}$. For
each $\rd\in\Kr(\M)$ containing $p_1$,  we can find
$\rd_1\in\Kr(\M)$ such that $p_1\in\rd_1\subset \rd$
and $\rd_1\perp\{p\}$.
This and $(*)$ imply that
$\mathfrak{R}_p \subset \mathfrak{R}_{p_1}$ and, by the symmetry of
$\perp$,
that $\mathfrak{R}_p =\mathfrak{R}_{p_1}$. For a generic $q\in\M$
we can find a finite sequence $p_1,\ldots, p_n$,
such that $\{p\}\perp \{p_1\}$,
$\{p_1\}\perp \{p_2\},\ldots,\{p_n\}\perp \{q\}$. Consequently,
$\mathfrak{R}_p=\mathfrak{R}_{q}$ and
$\mathfrak{R}_p\subset\pi(\A(\rd))''\cap\pi(\A(\rd))'$  for each
$\rd\in\Kr(\M)$. We also have that
$\mathfrak{R}_p\subset\pi(\A(\dc))''\cap\pi(\A(\dc))'$
for any $\dc\in\K(\M)$ such that $\M\setminus\overline{\dc}$ is nonempty.
The irreducibility of $\pi$ and (\ref{Cb:0}) complete the proof.
\end{proof}
For later purposes, it is useful to note that  if
$(\Al_{\K(\M)},\omega)$ satisfies Haag duality, then
$(\Al_{\K(\M)},\omega)$ is \textit{outer regular}, that is for any
$\rd_1\in\Kr(\M)$ we have
\begin{equation}
\label{Cb:2}
\pi(\A(\rd_1))''  =  \cap \ \{\pi(\A(\rd))'' \ | \
    \ \overline{\rd}_1\subset \rd\in\Kr(\M) \}.
\end{equation}
In fact, consider $\rd_2\in\Kr(\M)$ such that
$\rd_2\perp \rd_1$. This means that
$\overline{\rd}_1$ is contained in the open set
$\M\setminus\J(\overline{\rd}_2)= \M\setminus\overline{\J(\rd_2)}$.
By (\ref{Ba:3})
there is a regular diamond $\rd$ such that
$\overline{\rd}_1\subset \rd$ and
$\rd\perp\rd_2$. This  with the
observation that $\pi(\A(\rd))''\subset \pi(\A(\rd_2))'$, imply that
\begin{multline*}
\pi(\A(\rd_1))'' \ \subseteq \
\cap \{ \ \pi(\A(\rd))'' \ | \rd\in\Kr(\M), \
\overline{\rd}_1\subset \rd \} \\
\subseteq \
\cap \{\pi(\A(\rd_2))' \ | \rd_2\in\Kr(\M), \ \rd_2 \perp \rd_1\}
 = \pi(\A(\rd_1))''
\end{multline*}
completing the proof of (\ref{Cb:2}).
\subsection{Punctured Haag duality in a locally covariant quantum 
             field theory}
\label{Cc}
In the Haag-Kastler framework we have seen that punctured Haag
duality implies Haag duality and local definiteness. 
On the other hand, in the setting of the locally covariant 
quantum field theories, the relation between these properties 
is stronger. To make a precise claim, let us consider  
a locally covariant quantum field theory $\Al$, and let 
$\mathfrak{S}_{\mathrm{HL}}(\Al)$, $\mathfrak{S}_{\mathrm{pH}}(\Al)$ 
be two sets of state spaces of $\Al$ defined as follows:
\begin{itemize}
\item $\mathfrak{S}_{\mathrm{HL}}(\Al)$  is the family of state spaces $\St$ of
$\Al$ such that, for any $\M\in\Man$  and  for any  pure state 
	 $\omega\in\St(\M)$ the pair 
         $(\Al_{\K(\M)},\omega)$ is locally definite and satisfies 
         Haag duality;
\item $\mathfrak{S}_{\mathrm{pH}}(\Al)$ is the family of state spaces $\St$ of
$\Al$ such that,  for any $\M\in\Man$  and  for any  pure state 
	 $\omega\in\St(\M)$ the pair  $(\Al_{\K(\M)},\omega)$   
         satisfies punctured Haag duality and (\ref{Cb:0}). 
\end{itemize}
Then, we will  prove that 
$\mathfrak{S}_{\mathrm{HL}}(\Al)= \mathfrak{S}_{\mathrm{pH}}(\Al)$. 
Notice that 
$\mathfrak{S}_{\mathrm{pH}}(\Al)\subseteq\mathfrak{S}_{\mathrm{HL}}(\Al)$ 
because of Proposition \ref{Cb:1}. Moreover, one can easily see that,
if $\St\in\mathfrak{S}_{\mathrm{HL}}(\Al)$ and $\omega\in\St(\M)$,
then $\omega$ is a pure state of $\Al(\M)$ and 
$(\Al_{\K(\M)},\omega)$ fulfils (\ref{Cb:0}).
So, what remains to prove is that $(\Al_{\K(\M)},\omega)$
satisfies punctured Haag duality.\\[5pt] 
\indent To begin with, let us take $\St\in\mathfrak{S}_{\mathrm{HL}}(\Al)$, 
$\M\in\Man$, $p\in\M$,  and a  state $\omega\in\St(\M)$. 
Let $\M_p$ be the  causal excision of $p$, and let 
\[
\Al_{\K(\M_p)}: \K(\M_p)\ni\dc\longrightarrow \A_p(\dc)\subseteq
\Al(\M_p)
\]
be the net associated with $\M_p$, where the subscript $p$ is added
in order to avoid confusion  between
the elements of $\Al_{\K(\M_p)}$ and those of $\Al_{\K(\M)}$.
Observing that
$\io_p\in\Hom(\M_p,\M)$,  we can define
\[
\omega_p(A) \defi \al_{\io_p}^*\omega(A) =  \omega(\al_{\io_p}(A)),
\qquad  A\in\Al(\M_p).
\]
Since $\omega\in\St(\M)$ and
$\omega_p=\alpha^*_{\io_p}\omega$, by the definition of state space
$\omega_p$ belongs to $\St(\M_p)$. 
The first step of our proof consists in showing that
$(\Al_{\K(\M_p)},\omega_p)$ satisfies Haag
duality that, according to the definition of $\mathfrak{S}_{\mathrm{HL}}(\Al)$
it is equivalent to prove that $\omega_p$ is pure. To this aim let
us define
\[
V_{\io_p} \pi_p(A) \Omega_p \defi \pi(\alpha_{\io_p}(A)) \Omega,
\qquad A\in \Al(\M_p),
\]
where $(\Hil, \pi,\Omega)$ and $(\Hil_p, \pi_p,\Omega_p)$ are
respectively the GNS constructions associated with $\omega$ and $\omega_p$.
\begin{prop}
\label{Cc:2}
$(\Al_{\K(\M_p)},\omega_p)$  satisfies Haag duality.
In particular:\\
a) the representation $\pi\circ\alpha_{\io_p}$ of $\Al(\M_p)$
is irreducible; \\
b) $V_{\io_p}\in (\pi_p,\pi\circ\alpha_{\io_p})$ is unitary,
hence $\omega_p$ is pure.
\end{prop}
\begin{proof}
a) Let $T\in (\pi\circ\alpha_{\io_p},\pi\circ \alpha_{\io_p})$.
Given $\rd_1\in\Kr(\M)$ such that $p\in \rd_1$,  let
us consider $\rd\in\Kr(\M), \ \rd\perp \rd_1$. By (\ref{Bb:5})
$\rd$ belongs to $\K(\M_p)$ and  by (\ref{Ca:1})
$\A(\rd) = \alpha_{\io_p}(\A_p(\rd))$.
Thus, $T\in (\pi\upharpoonright \A(\rd),\pi\upharpoonright \A(\rd))$
for each regular diamond of
$\M$ causally disjoint from $\rd_1$. By Haag duality
$T\in \pi(\A(\rd_1))''$ for each $\rd_1\in\Kr(\M)$ such that $p\in\rd_1$;
hence by local definiteness $T=c\cdot \mathbbm{1}$ completing the proof.
b) Observe that for each
$A\in\Al(\M_p)$ we have
$\norm{V_{\io_p}\pi_p(A)\Omega_p}^2$
$=\norm{\pi(\alpha_{\io_p}(A))\Omega}^2$
$= (\Omega, \pi(\alpha_{\io_p}(A^*A))\Omega)$
$= \omega_p(A^*A) = \norm{\pi_p(A)\Omega_p}^2$. This entails that
$V_{\io_p}$ is a unitary intertwiner between
$\pi_p$ and $\pi\circ\alpha_{\io_p}$ because
$\Omega_p$ is cyclic for $\pi_p$ and $\pi\circ\alpha_{\io_p}$ is irreducible.
Therefore $\pi_p$ is irreducible and, consequently, $\omega_p$ is
pure. Finally, as observed above, $(\Al_{\K(\M_p)},\omega_p)$
satisfies Haag duality.
\end{proof}
We do not know whether the sets $\Kr(\M_p)$ and $\{\rd\in\Kr(\M) \ | \
\rd\perp\{p\}\}$ are equal or not. If they were the same,
punctured Haag duality for $(\Al_{\K(\M)},\omega)$ would follow from
Haag duality for $(\Al_{\K(\M_p)},\omega_p)$. Nevertheless,
by Proposition \ref{Bb:7} the sets
$\Kr(\M_p)$ and $\{\rd\in\Kr(\M) \ | \ \rd\perp\{p\}\}$
have a common ``dense'' subset: the set  $\Kr(\M_p\wedge\M)$. 
This is enough for our aim and
will allow us to prove punctured Haag duality in two steps.
\begin{lemma}
\label{Cc:3}
For any $\rd_1\in\Kr(\M_p\wedge \M)$ the following identity holds
\[
\pi(\A(\rd_1))''  = \cap \{ \ \pi(\A(\rd))' \ | \ \rd\in\Kr(\M), \
\rd\perp  \left(\rd_1\cup \{p\}\right)  \}
\]
\end{lemma}
\begin{proof}
Let us consider $\rd_2\in\Kr(\M_p)$,  $\rd_2\perp\rd_1$. The closure
of $\rd_2$ is contained in the open set
$\M\setminus\J(\overline{\rd}_1 \cup \{p\})$
(see Section \ref{A}). By Proposition~\ref{Bb:7} 
there is $\rd_o\in\Kr(\M_p \wedge\M)$ satisfying the relations
$\overline{\rd}_2 \subset \rd_o$,
$\rd_o\perp (\rd_1\cup \{p\})$.
This leads to the following inclusions
\begin{multline*}
\pi(\A(\rd_1))'' \subseteq  \ \cap \
  \{ \ \pi(\A(\rd_o))' \ | \ \rd_o\in\Kr(\M), \
              \rd_o \perp (\rd_1\cup \{p\}) \} \\
    \subseteq
\cap \ \{ \ \pi(\A(\rd_2))' \ | \rd_2\in\Kr(\M_p), \
             \rd_2\perp\rd_1\} \ \ \ (*)
\end{multline*}
Recall now that by  Proposition \ref{Cc:2}.b
$\pi(\A(\rd_1))'' = V_{\io_p}\pi_p(\A_p(\rd_1))''V^*_{\io_p}$.
As $(\Al_{\K(\M_p)},\omega_p)$ satisfies Haag duality we have
\begin{multline*}
 \pi(\A(\rd_1))''
     = V_{\io_p} \left( \cap \{ \ \pi_p(\A_p(\rd_2))'  
        \ |\rd_2\in\Kr(\M_p), \
          \rd_2 \perp \rd_1\}\right)  V^*_{\io_p} \\
    =  \cap \{ \ \pi(\A(\rd_2))' \ | \ \rd_2\in\Kr(\M_p), \
               \rd_2\perp\rd_1\}
\end{multline*}
Combining this with $(*)$ we obtain the proof.
\end{proof}
\begin{teo}
\label{Cc:4}
Given $\St\in\mathfrak{S}_{\mathrm{HL}}(\Al)$, for any $\M\in\Man$ and
for any $\omega\in\St(\M)$ the pair 
$(\Al_{\K(\M)},\omega)$ satisfies punctured Haag duality. 
\end{teo}
\begin{proof}
Let $\omega\in\St(\M)$ be a pure state of $\Al(\M)$, and let 
$\pi$ be the GNS representation associated with $\omega$. 
Fix $p\in\M$ and $\rd_1\in\Kr(\M)$ such that $\rd_1\perp\{p\}$.
Notice that if  $\rd_2\in\Kr(\M)$ such that $\overline{\rd}_1\subset
\rd_2$, then  $\overline{\rd}_1$ is contained in the open set 
$\rd_2\cap (\M\setminus\J(p))$. By Proposition \ref{Bb:7}
there is $\rd_o\in\Kr(\M_p\wedge\M)$ such that
$\overline{\rd}_1\subset \rd_o$,
$\overline{\rd}_o\subset \rd_2$ and  $\rd_o\perp \{p\}$.
As $\rd_o$ fulfils the hypotheses of  Lemma
\ref{Cc:3}, we have
\begin{align*}
\pi(\A(\rd_1))'' \ \subseteq &  \ \cap \ \{ \ \pi(\A(\rd))' \ | \rd\in\Kr(\M), \  \rd \perp
     (\rd_1\cup \{p\}) \}\\
\subseteq  &
 \ \cap \ \{\pi(\A(\rd))' \ | \rd\in\Kr(\M), \
      \rd\perp(\rd_o\cup \{p\}) \}\\
 = & \ \pi(\A(\rd_o))''\subseteq \pi(\A(\rd_2))''.
\end{align*}
These inclusions are verified for any $\rd_2\in\Kr(\M)$ such that
$\overline{\rd}_1\subset \rd_2$.  Hence, the outer
regularity (\ref{Cb:2}) implies that $(\Al_{\Kr(\M)},\omega)$ 
satisfies punctured Haag duality. 
\end{proof}
This theorem and 
the observations at the beginning of this section lead to the following
\begin{cor}
\label{Cc:5}
$\mathfrak{S}_{\mathrm{HL}}(\Al)=\mathfrak{S}_{\mathrm{pH}}(\Al)$. 
\end{cor}
\subsection{The case of the free Klein-Gordon field}
\label{Cd}
The theory of the free  Klein-Gordon field provides, as shown in \cite{BFV},
an example of a locally covariant quantum
field theory $\mathscr{W}$ with a state space $\St_\mu$.
The functor $\mathscr{W}$ is defined as the correspondence
$\M\longrightarrow\mathscr{W}(\M)$ that associates to
any $\M\in\Man$ the Weyl algebra $\mathscr{W}(\M)$ of the free
Klein-Gordon field over $\M$.  $\St_{\mu}$ is defined
as the correspondence $\M\longrightarrow\St_{\mu}(\M)$
that to any $\M$ associates  the collection of
the states of $\mathscr{W}(\M)$ which are
locally quasiequivalent
to quasi-free states of $\mathscr{W}(\M)$  fulfilling
the microlocal spectrum condition (or equivalently the Hadamard
condition). We refer the reader to the cited paper and reference therein
for a detailed description  of this example.
We now prove that for any $\omega\in\St_\mu(\M)$ the pair 
$(\mathscr{W}_{\K(\M)}, \omega)$ is locally definite and, if $\omega$ is 
pure, satisfies Haag duality. Hence,  
$\St_\mu\in\mathfrak{S}_{\mathrm{HL}}(\mathscr{W})$,  and 
by Theorem \ref{Cc:4} 
$(\mathscr{W}_{\K(\M)}, \omega)$ satisfies punctured Haag duality 
for any pure state $\omega\in\St_\mu(\M)$.\\[3pt]
\indent Let us start by recalling that two states $\omega,\omega_1$ of
$\mathscr{W}(\M)$ are said to be  \textit{locally quasiequivalent}
if for each $\dc\in\K(\M)$ there exists
an isomorphism $\rho_\dc:\pi(\mathcal{W}(\dc))''\longrightarrow
\pi_1(\mathcal{W}(\dc))''$  such that
$\rho_\dc\pi(A)=\pi_1(A)$ for each $A\in\mathcal{W}(\dc)$, where
$\pi,\pi_1$ are respectively the GNS representations of $\omega$ and
$\omega_1$.
Furthermore, we need to recall the following fact
(\cite[Theorem 3.6]{Ve}): for each
$\M\in\mathfrak{Man}$,
if $\omega_\mu$ is a quasi-free state
of $\mathscr{W}(\M)$ satisfying the microlocal spectrum condition,
then $(\mathscr{W}_{\K(\M)},\omega_\mu)$ is locally definite and,
if $\omega_\mu$ is pure, it satisfies Haag duality.
\begin{prop}
\label{Cd:1}
$\St_{\mu}\in\mathfrak{S}_{\mathrm{HL}}(\mathscr{W})$. Therefore,
$(\mathscr{W}_{\K(\M)}, \omega)$ satisfies punctured Haag duality
for any pure state $\omega\in\St_{\mu}(\M)$.
\end{prop}
\begin{proof}
Fix  $\M\in\Man$  and
consider a pure state $\omega$ of $\mathscr{W}(\M)$ which is
locally quasiequivalent to a quasi-free state $\omega_\mu$
satisfying the microlocal spectrum condition. Let $\pi$ and $\pi_\mu$
be the GNS representations of $\omega$ and $\omega_\mu$ respectively.
It has  already been shown in \cite{Ve} that
$(\mathscr{W}_{\K(\M)}, \omega)$ satisfies Haag duality.
Hence it remains to be proved that
$(\mathscr{W}_{\K(\M)}, \omega)$ is locally definite.
To this aim,  fix
$p\in\M$ and $\rd_1\in\Kr(\M)$ such that $p\in\rd_1$. As observed above
$(\mathscr{W}_{\K(\M)},\omega_\mu)$ is locally definite. This entails that
$\cap \{\pi_\mu(\mathcal{W}(\rd))'' \ |\rd\in\Kr(\M), \ p\in\rd\subset\rd_1\}
=\mathbb{C}\cdot \mathbbm{1}$
because $\Kr(\M)$ contains a neighbourhoods basis of $p$.
Being $\omega$ locally  quasiequivalent to $\omega_\mu$, there is
an isomorphism $\rho_{\sst{\rd_1}}$ from $\pi(\mathcal{W}(\rd_1))''$ onto
$\pi_\mu(\mathcal{W}(\rd_1))''$
such that $\rho_{\sst{\rd_1}}\pi(A)=\pi_\mu(A)$ for each
$A\in\mathcal{W}(\rd_1)$. Hence
\[
\rho_{\sst{\rd_1}}:
\cap \{ \ \pi(\mathcal{W}(\rd))'' \ |p\in\rd\subset\rd_1\}
\longrightarrow
\cap \{ \ \pi_\mu(\mathcal{W}(\rd))'' \ |p\in\rd\subset\rd_1\}
=\mathbb{C}\cdot \mathbbm{1}
\]
is an isomorphism and, consequently,
$(\mathscr{W}_{\K(\M)},\omega)$ is locally definite.
\end{proof}
\begin{oss}
\label{Cd:2}
It is worth mentioning that the family of  the adiabatic vacuum states
of order $N>\frac{5}{2}$, studied in \cite{JS}, is contained in $\St_\mu$:
any such state is locally quasiequivalent to a
quasi-free state fulfilling the microlocal spectrum
condition.
\end{oss}
\appendix
\numberwithin{equation}{section}
\section{Proof of the Proposition \ref{Bb:7}}
\label{ApA}
The proof of the Proposition \ref{Bb:7} comes by
a slight modification of the Lemmas 5, 7 and 8 of \cite{Ve1}.
So, we give a detailed description only of the modified parts
of the proofs and  refer the reader to the cited paper for the
assertions that we will not prove.\\[5pt]
\indent We start by recalling some  results of the cited paper.
Let $\M\in\Man$ and let
$F:\R\times \Sigma \longrightarrow \M$ be a foliation of $\M$
by spacelike  Cauchy surfaces. For each  acausal (spacelike) Cauchy
surface $\C$, there is an associated pair
$(\tau_{\sst{\C}}, \f_{\sst{\C}})$ where
$\tau_{\sst{\C}}:
\Sigma\longrightarrow\R$ is a continuous (smooth) function, while
$\f_{\sst{\C}}$, defined as
\[
\f_{\sst{\C}}(y)\defi F(\tau_{\sst{\C}}(y),y), \qquad y\in\Sigma,
\]
is an homeomorphism (diffeomorphism) $\f_{\sst{\C}}:\Sigma\longrightarrow\C$.
Given another acausal Cauchy
surface $\C_o$ and the  corresponding
pair $(\tau_{\sst{\C_o}}, \f_{\sst{\C_o}})$, the map
$\Phi_{\sst{\C_o,\C}}:
\C\longrightarrow \C_o$
defined as
\[
\Phi_{\sst{\C_o,\C}}(p) \defi (\f_{\sst{\C_o}}\circ \f^{-1}_{\sst{\C}}) (p),
\qquad \forall p\in\C,
\]
is an homeomorphism (diffeomorphism if $\C$ and $\C_o$ are spacelike).\\[10pt]
\indent Now, consider the causal excision  $\M_p$  of $p\in\M$, and 
a spacelike Cauchy surface $\C_p$ 
of $\M_p$. By Proposition
\ref{Bb:3} $\C \defi \C_p\cup\{p\}$ is an acausal Cauchy surface of $\M$.
\begin{lemma}
\label{ApA:1}
For each continuous strictly
positive function $\eps: \Sigma\longrightarrow \R$ there exists
a spacelike Cauchy surface $\C_o$ that meets $p$, such that
$|\tau_{\sst{\C}}(y) -\tau_{\sst{\C_o}}(y)|<\eps(y)$
for each $y\in\Sigma$.
\end{lemma}
\begin{proof}
Let us define
$A^\pm_{\la} \defi \{ F(\tau_{\sst{\C}}(y)\pm\la\cdot \eps(y), y) \ | \ y\in
 \Sigma \}$,  $0<\la<1$, and
$\mathcal{N}\defi
\left(\M\setminus \left(\J^+(A^+_\la) \cup \J^-(A^-_\la)\right)\right)^o$.
Notice that
if $p=F(t_o,y_o)$ with $t_o >  \tau_{\sst{\C}}(y_o)+\la\cdot \eps(y_o)$,
then $p\not\in \mathcal{N}$.
In fact, the f-d timelike curve
\[
\ga(t)\defi F(\tau_{\sst{\C}}(y_o)+ t\cdot \eps(y_o),y_o),\qquad t\in \ [\la \ , \
 (t_o -\tau_{\sst{\C}}(y_o))\cdot {\eps(y_o)}^{-1}],
\]
joins the point $F(\tau_{\sst{\C}}(y_o)+ \la\cdot \eps(y_o),y_o)\in
A^+_{\la}$ with $p$.
Analogously, if  $p=F(t_o,y_o)$ with $t_o <  \tau_{\sst{\C}}(y_o)-\la\cdot
\eps(y_o)$,
then $p$ does not belong to $\mathcal{N}$. Hence
\[
p\in \mathcal{N}, \  p=F(t,y)  \ \ \iff  \ \ |t-\tau_{\sst{\C}}(y)| < \la\cdot \eps(y) \ \
 \Rightarrow  \ \ |t-\tau_{\sst{\C}}(y)|< \eps(y).
\]
Now, as $\mathcal{N}$
is globally hyperbolic,  there is
a spacelike Cauchy surface $\C_o$ of $\mathcal{N}$ that meets
$p$. $\C_o$ is also a spacelike Cauchy surface of $\M$.
Since $\C_o\subset \mathcal{N}$, we have
$|\tau_{\sst{\C}}(y) -  \tau_{\sst{\C_o}}(y)| < \eps(y)$
for each  $y\in\Sigma$
\end{proof}
\begin{lemma}
\label{ApA:2}
Let $\C_p$ and $\C$ as above.
Consider three connected, relatively compact,
open subsets  $G, U_1,U_2$  of $\C_p$  that verify
$\overline{G} \subset U_1$,
$\overline{U}_1 \subset U_2$.
Then, there exists a smooth acausal Cauchy surface $\C_o$ of $\M$
that meets $p$,such that:
\[
a) \ \J\left(\overline{G}\right)\cap\C_o \subset \Phi_{\sst{\C_o,\C}}(U_1), \qquad
b) \ \J\left(\overline{\Phi_{\sst{\C_o,\C}}(U_1)}\right) \cap \C \subset U_2.
\]
\end{lemma}
\begin{proof}
The sets $\mathcal{N}_{1\pm}$, defined as  $(\D^\pm(U_1))^o$,
are globally hyperbolic. Let us take
two spacelike Cauchy surfaces   $\St_{1\pm}$ of $\mathcal{N}_{1\pm}$.
Notice that $\C_{1\pm} \defi \St_{1\pm} \cup (\C\setminus U_1)$
are acausal Cauchy surfaces of $\M$, and that,
there are two strictly positive continuous functions
$\eps_{1\pm}: \f^{\sst{-1}}_{\sst{\C}}(U_1)\longrightarrow \R$ such that
$\St_{1\pm} = \{ F(\tau_{\sst{\C}}(y)\pm \eps_{1\pm}(y),y) \ | \ y\in
    \f^{\sst{-1}}_{\sst{\C}}(U_1)\}$.
Let us define
\[
U_{1+}\defi \J^-(\J^+(\overline{G})\cap \St_{1+}) \cap \C \ \ , \ \
U_{1-}\defi \J^+(\J^-(\overline{G})\cap \St_{1-}) \cap \C.
\]
Since $\J^+(\overline{G})\cap \C_{1+}$ is a closed subset of
$\C_{1+}$ and
$\J^+(\overline{G})\cap \C_{1+} = \J^+(\overline{G})\cap \St_{1+}$,
we have that $U_{1+}$ is a closed subset of $U_1$ and
$G\subset U_{1+}$. The same holds for $U_{1-}$.
The set $W_1\defi U_{1+}\cup U_{1-}$
is closed, compact
(because contained in a relatively compact set)
and $G \subset W_1 \subset U_1$. We now apply the same reasoning
with respect to the inclusion
$U_1\subset U_2$. Namely, given two spacelike Cauchy surfaces
$\St_{2\pm}$ of the spacetimes
$(\D^\pm(U_2))^o$,  we consider  the acausal Cauchy surfaces $\C_{2\pm}$ of $\M$
defined as $\C_{2\pm}\defi \St_{2\pm} \cup (\C\setminus U_2$).
As above, there are two
strictly positive continuous functions $\eps_{2\pm}$ such that
$\St_{2\pm} = \{ F(\tau_{\sst{\C}}(y)\pm\eps_{2\pm}(y),y) \ | \ y\in
    \f^{\sst{-1}}_{\sst{\C}}(U_2)\}$.
Thus, we can find a compact set $W_2$ of $\C$ verifying
$\overline{U}_1\subset W_2\subset U_2$. Now let us define
\[
\eps \defi \min \left\{\min_{y\in \f^{\sst{-1}}_{\sst{\C}}(W_1)}
 \{\eps_{1+}(y),\eps_{1-}(y)\} \ , \
  \min_{y\in \f^{\sst{-1}}_{\sst{\C}}(W_2)}
   \{\eps_{2+}(y),\eps_{2-}(y)\}\right\}.
\]
By Lemma \ref{ApA:1} there is a spacelike Cauchy surface $\C_o$ of $\M$
that meets $p$, such that
$|\tau_{\sst{\C_o}}(y) - \tau_{\sst{\C}}(y) | < \eps$  for each
$y\in\Sigma$. Since $\overline{G}\subseteq W_1$,
by the definition of $\eps$
the set $\J^+(\overline{G})\cap \C_o$ is in the past of
$\J^+(\overline{G})\cap S_{1+}$, while
$\J^-(\overline{G})\cap \C_o$ is in
the future of $\J^-(\overline{G})\cap S_{1-}$. Hence, we have
\[
\J\left(\J(\overline{G}) \cap \C_o\right)  \cap \C\subset W_1.
\]
This entails
\begin{align*}
\Phi_{\sst{\C,\C_o}}(\J(\overline{G}) \cap \C_o) \ \  = \ \ \ & \ \
\f_{\sst{\C}}\circ \f^{\sst{-1}}_{\sst{\C_o}}(\J(\overline{G}) \cap \C_o)
\subset W_1  \\
  \iff & \J(\overline{G}) \cap \C_o \subset
       \Phi_{\sst{\C_o,\C}}(W_1) \ \
 \Rightarrow \ \   \J(\overline{G}) \cap \C_o \subset
\Phi_{\sst{\C_o,\C}}(U_1),
\end{align*}
completing the proof of the statement a). The same reasoning
applied to the inclusion $U_1\subset U_2$ leads to
$\J(\J(\overline{U}_1)\cap \C_o)\cap \C \subset W_2$.
As $\Phi_{\sst{\C_o,\C}}(U_1)$ is contained in the closed set
$\J(\overline{U}_1)\cap \C_o$, we have that
$\J(\overline{\Phi_{\sst{\C_o,\C}}(U_1)}) \cap \C \subset W_2\subset U_2$.
\end{proof}
\begin{prop}
\label{ApA:3}
Let $\rd\in\Kr(\M_p)$ and let $V$ be an open set of $\M_p$
such that $\overline{\rd}\subset V$.
Then, there exist a spacelike Cauchy surface
$\C_o$ of $\M$ that meets $p$, and
$\rd_o\in\Kr(\M,\C_o)$ such that
$\overline{\rd}\subset \rd_o$, and   $\overline{\rd}_o\subset(\D(V))^o$.
\end{prop}
\begin{proof}
Assume that $\rd$ is based on a spacelike Cauchy surface
$\C_p$ of $\M_p$ and let $G\subset \C_p$ be the base of $\rd$.
The set $U\defi V\cap \C_p$
is open in $\C_p$ and $\overline{G}\subset U$. According to \cite[Lemma 3]{Ve1}
we can find $U_1,U_2\in\mathfrak{G}(\C_p)$ such that
$\overline{G}\subset U_1$,  $\overline{U}_1\subset U_2$ and
$\overline{U}_2\subset U$.
Let $\C$ be the acausal Cauchy surface of $\M$ defined as $\C_p\cup \{p\}$.
By  Lemma \ref{ApA:2}  there is a spacelike Cauchy surface $\C_o$
of $\M$ that meets $p$, such that
\[
\J\left(\overline{G}\right)\cap\C_o \subset \Phi_{\sst{\C_o,\C}}(U_1), \qquad
\J\left(\overline{\Phi_{\sst{\C_o,\C}}(U_1)}\right) \cap \C \subset U_2.
\]
$\f_{\sst{\C_o}}$ is a diffeomorphism
between $\Sigma$ and $\C_o$ because $\C_o$ is spacelike.
Notice now  that $\f_{\sst{\C}}: \Sigma \longrightarrow  \C_p\cup \{p\}$ is an
homeomorphism because $\C =\C_p\cup \{p\}$ is an acausal,
in general nonsmooth, Cauchy surface of $\M$.
However,  $\C_p$ is  spacelike, hence smooth. Then,  it easily follows
from the definition
of $\f_{\sst{\C}}$  that
$\f_{\sst{\C}}: \Sigma\setminus \{\f^{\sst{-1}}_{\sst{\C}}(p) \}
\longrightarrow  \C_p$ is a diffeomorphism.
Then,  $\rd_o\defi(\D(\Phi_{\sst{\C_o,\C}}(U_1)))^o$ is
a regular diamond of $\M$ based on $\C_o$.
The previous inclusions entail that
$\overline{\rd}\subset\rd_o$,
$\overline{\rd}_o \subset (\D(U_2))^o\subset(\D(V))^o$
completing the proof.
\end{proof}
Because  (\ref{Bb:6}) the Proposition \ref{ApA:3} proves the first part of the 
Proposition \ref{Bb:7}. 
Concerning the second part, we have the following
\begin{prop}
\label{ApA:4}
Let $\rd\in\Kr(\M)$  be such that $\rd\perp\{p\}$ and
let $V$ be an open set of $\M$
such that $\overline{\rd}\subset V$.
Then, there exist a spacelike Cauchy surface
$\C_o$ of $\M$, that meets $p$, and
$\rd_o\in\Kr(\M,\C_o)$ such that
$\overline{\rd}\subset \rd_o$, $\overline{\rd}_o\subset(\D(V))^o$ and
$\rd_o\perp\{p\}$.
\end{prop}
\begin{proof}
The proof is very similar to the proof of Proposition \ref{ApA:3}.
Assume that $\rd=(\D(G))^o$ where $G\in\mathfrak{G}(\C)$ for some
spacelike Cauchy surface $\C$ of $\M$.
Notice $\overline{G}\cap\J(p)=\emptyset$.
Let us define $U=(\C\cap V) \setminus (\C\cap\J(p))$.
$U$ is an open set of $\C\setminus (\C\cap\J(p))$ and
$\overline{G}\subset U$. By \cite[Lemma 3]{Ve1}  we can find
$U_1,U_2\in\mathfrak{G}(\C)$ such that $\overline{G}\subset U_1$,
$\overline{U}_1\subset U_2$ and
$\overline{U}_2\subset U$. Now, notice that in general $\C$ does not
meet $p$, hence the Lemmas \ref{ApA:1} and
\ref{ApA:2} cannot be applied.
In this case, however, we can use directly \cite[Lemma 5]{Ve1}
that asserts that for each $\eps>0$ there exists a
spacelike Cauchy surface $\C_o$ that
meets $p$  such that
$|\tau_{\sst{C}}(y)-\tau_{\sst{\C_o}}(y)|<\eps$ for any
$y\in \f^{\sst{-1}}_{\sst{\C}}(U)$.  Proceeding as in Lemma \ref{ApA:2},
one can choose $\eps$  in such a way that
$\J\left(\overline{G}\right)\cap\C_o \subset \Phi_{\sst{\C_o,\C}}(U_1)$, 
$\J\left(\overline{\Phi_{\sst{\C_o,\C}}(U_1)}\right) \cap \C \subset U_2$.
Proceeding now as in Proposition \ref{ApA:3},  these inclusions lead to the
proof of the statement.
\end{proof}
\noindent {\small \textbf{Acknowledgements.}
I would like to thank  Daniele Guido and John E. Roberts
for helpful discussions and their constant interest in this work.
Finally, I am grateful to Gerardo Morsella for his precious help.}

\end{document}